\documentclass[twocolumn]{aastex63}
\usepackage{color}

\received{December 19, 2019}
\revised{January 16, 2020}
\accepted{February 2, 2020}

\submitjournal{ApJL}

\shorttitle{Blue transients at the high-z universe}
\shortauthors{Terasaki et al.}

\begin{document}

\title{Fast Luminous Blue Transients in the Reionization Era and Beyond}

\correspondingauthor{Terasaki Tomoki}
\email{tterasaki@cmb.phys.s.u-tokyo.ac.jp}

\author{Tomoki Terasaki}
\affiliation{Department of Astronomy, School of Science, The University of Tokyo, 7-3-1 Hongo, Bunkyo-ku, Tokyo 113-0033, Japan}

\author{Daichi Tsuna}
\affiliation{Research Center for the Early Universe (RESCEU), Graduate School of Science, The University of Tokyo, 7-3-1 Hongo, Bunkyo-ku, Tokyo 113-0033, Japan}
\affiliation{Department of Physics, Graduate School of Science, The University of Tokyo, 7-3-1 Hongo, Bunkyo-ku, Tokyo 113-0033, Japan}

\author{Toshikazu Shigeyama}
\affiliation{Research Center for the Early Universe (RESCEU), Graduate School of Science, The University of Tokyo, 7-3-1 Hongo, Bunkyo-ku, Tokyo 113-0033, Japan}
\affiliation{Department of Astronomy, School of Science, The University of Tokyo, 7-3-1 Hongo, Bunkyo-ku, Tokyo 113-0033, Japan}

\begin{abstract}
To determine the epoch of reionization precisely and to reveal the property of inhomogeneous reionization are some of the most important topics of modern cosmology.
Existing methods to investigate reionization which use cosmic microwave background, Ly$\alpha$ emitters, quasars, or gamma ray bursts, have difficulties in terms of accuracy or event rate.
We propose that recently discovered fast luminous blue transients (FLBTs) have potential as a novel probe of reionization.
We study the detectability of FLBTs at the epoch of reionization with upcoming WFIRST Wide-Field Instruments (WFI), using a star formation rate derived from galaxy observations and an event rate of FLBTs proportional to the star formation rate.
We find that if FLBTs occur at a rate of 1\% of the core-collapse supernova rate,  2 (0.3) FLBTs per year per deg$^2$ at $z>6$ ($z>8$) can be detected by a survey with a limiting magnitude of 26.5 mag in the near-infrared band and a cadence of 10 days.
We conclude that the WFIRST supernova  deep survey can detect $\sim20$ FLBTs at the epoch of reionization in the near future.

\end{abstract}


\section{Introduction} \label{sec:intro}
Observations of the cosmic microwave background (CMB) indicate that at redshift $z\sim1100$, electrons were captured by nuclei and the universe became neutral \citep{Spergel2003}. 
On the other hand, the present-day universe is ionized, with the intergalactic medium (IGM) dominated by a plasma of hot electrons.
Therefore, reionization of the IGM should have occurred in the early universe. 

The precise identification of the history and source(s) of reionization is one of the important topics in modern cosmology.
Recently the optical depth of Thomson scattering was obtained from Planck CMB data to be $\tau\sim0.06$, and the epoch of reionization was constrained in the range $z_{\rm{re}}\sim 7$--$9$ \citep{Planck16reionization,Planck2018,Pagano19}, which roughly supports the values of WMAP \citep{WMAP2013}.
However, this estimation relies on modelling of the time evolution of the ionization fraction, and the precise evolution is difficult to pin down just from an integrated value.

A different approach using individual objects has been historically done using absorption of Ly$\alpha$ photons by neutral hydrogen. Because of the large cross section of the Ly$\alpha$ absorption and the effect of the cosmic expansion, most of photons with energies higher than the Ly$\alpha$ line are absorbed by neutral hydrogen. This feature is called the Gunn-Peterson trough \citep{Gunn1965}, and the signature of this trough sets a lower limit on the fraction $x_{\rm{HI}} \equiv n_{\rm{HI}}/n_{\rm{H}} \gtrsim 10^{-3}$ of neutral hydrogen in the IGM. 
Moreover, the fraction of neutral hydrogen in the IGM can be more precisely derived from the shape of a damping wing developing in the redder side of the Ly$\alpha$ absorption line \citep{Miralda-Escude1998}.
These physical features of Ly$\alpha$ absorption make bright objects at cosmological distances a probe for the ionization history of the universe.

One such probe is Ly$\alpha$ emitters (LAEs), which are galaxies that emit strong Ly$\alpha$ lines.
The number density of LAEs was found to decline suddenly beyond $z\gtrsim6$ (e.g. \citealt{Malhotra2004,Stark2011,Kashikawa2011}), which indicates a neutral IGM during that period.
However, this estimation of $x_{\rm{HI}}$ by the number density of LAEs is affected by the degree of galaxy clustering and tends to be underestimated, because a high degree of galaxy clustering can preferentially ionize these regions \citep{Dijkstra2014,Hu2019}. Therefore, to obtain information of the overall reionization, bright objects whose existence is independent of galaxy clustering are more helpful. 

Quasars and gamma-ray bursts (GRBs) are two bright candidates that can be observed at high redshifts. Quasars are the most luminous persistent sources and many quasars at high redshifts are known (e.g. \citealt{Fan2006,Becker2015,Bosman2018}).
However, there are some difficulties such as fluctuations of intrinsic spectra \citep{Barkana2004} and large HII regions, which weaken the effect of damping wing \citep{Madau2000}. Quasars are also expected to trace regions that are much more ionized than the global mean (e.g. \citealt{Alvarez07}). Thus measurement with quasars can lead to an underestimate of $x_{\rm{HI}}$ as the global mean (e.g. \citealt{Mesinger10}).

The afterglow of GRBs exhibit power-law spectra (e.g. \citealt{Sari98}), which are far simpler than the spectra of LAEs and quasars. This minimizes the uncertainty coming from spectral modelling.
The fraction of neutral hydrogen can be estimated by a simple red damping wing model, as was done for the afterglows of GRB050904 \citep{Totani2006} and GRB130606A \citep{Totani2013, Hartoog2015}.
However, a GRB can be observed only when an observer is located on the axis of the jet and the luminosity of their afterglow is high.
Thus, only a few GRBs at the epoch of the reionization have been observed to date and no studies of reionization have been done with GRB afterglows beyond $z\sim 6$.

In this work we suggest a novel probe for reionization, Fast Luminous Blue Transients (FLBTs). Recent high-cadence surveys have discovered many types of transients whose light curves, spectral energy distributions (SEDs), and line properties can not be explained by a standard supernova theory \citep{Drout2014, Tanaka2016, Pursiainen2018}. FLBTs are one of these newly discovered transients. The timescale of the transient is very short (half-decay time $t_{1/2}\lesssim10$ days), the luminosity is very high ($L_{\rm bol}\sim 10^{43}$--$10^{44}\ {\rm erg\ s^{-1}}$), and the temperature is very high ($T\gtrsim10,000\,\rm{K}$) as compared with the other known optical transients. Until the discovery of AT 2018cow, there were only distant examples ($z\gtrsim0.1$) and details of FLBTs were poorly known.

AT 2018cow is the first FLBT observed in the local universe at $z=0.0139$ \citep{ Prentice2018, Kuin09, Perley2019a}, which enabled multi-wavelength follow-up observations. The transient was very bright $L_{\rm{cow}} \sim 10^{10}\, L_{\odot}$ and rapid, with timescales of rising and declining are respectively $t_{\rm{rise},1/2} \sim 2.5\, \rm{days}$ and $t_{\rm{decline},1/2} \sim 3\, \rm{days}$. The origin of AT 2018cow is unknown, but from constraints of the line features, ejecta mass/velocity, and engine timescale, it is considered to be an explosion event that occurred in a dense circumstellar environment \citep{Margutti2019}. Because of the extremely high luminosity, the high temperature, and the spectral shape close to a Planck function, we can expect FLBTs as a good probe for reionization like GRBs, but observable at a comparable or higher rate. At high redshifts around the epoch of reionization, a substantial part of the emission from FLBTs are expected to be redshifted to near infrared (NIR) wavelengths, which are suitable for NIR transient surveys.

In this paper, we calculate the event rate of FLBTs and estimate their detectability with the upcoming WFIRST Wide-Field Instrument \citep{Spergel2015}. We also present an optimized observation strategy which maximizes the detection counts of FLBTs at the epoch of reionization.

We assume a flat $\Lambda$CDM model with $H_0 = 70\, \rm{km\,s^{-1}\,Mpc^{-1}}$, $\Omega _M = 0.3$, $\Omega _{\Lambda} = 0.7$.
All magnitudes are given in the AB magnitude system.

\section{Method} \label{sec:method}
To estimate the detectability of FLBTs, we construct a light curve model for an FLBT, AT 2018cow, and discuss its rate, the observation parameters, and the detection criteria in the following subsections. 

\subsection{Models of FLBTs}
We use the light curve of AT 2018cow as a representative model of FLBTs, since this is the most well-studied one. To construct the light curve we adopt the light curve data compiled in \citet{Perley2019a}.

The UV-optical spectral energy distribution (SED) of AT 2018cow is well fitted by a single Planck function, especially in the early days. As photometric data with wavelengths shorter than $200\,\rm{nm}$ are not available, we extrapolated the blackbody function to the FUV range. \cite{Perley2019a} pointed out that there is a non-thermal power-law component with a spectral index ($F_{\nu}\propto \nu^{\alpha}$) of $\alpha \sim -0.75$ in the NIR region. However, in the case of our interest of observing high-redshift ($z>5$) FLBTs in the NIR bands, the power-law component is completely outside the band. Therefore, it is sufficient for our estimation to use the black-body model and neglect the power-law component.

Neutral hydrogen gas along the line of sight should prevent UV photons with wavelengths shorter than the Ly$\alpha$ ($\lambda < 121.57(1+z)\,\rm{nm}$) from reaching us, especially when we observe a transient at high-redshift. Thus, we take a conservative approach and set the flux in these short-wavelength regions to be zero for all the redshifts.

From the $L_{\nu}$ spectrum thus obtained, we calculated the apparent AB magnitude by using Equation (8) of \citet{Hogg2002}.
We adopted the WFI filters of WFIRST \footnote{\url{https://wfirst.gsfc.nasa.gov/science/WFIRST_Reference_Information.html}} as the bandpass filters.

\begin{figure}
    \centering
    \includegraphics[width=\linewidth]{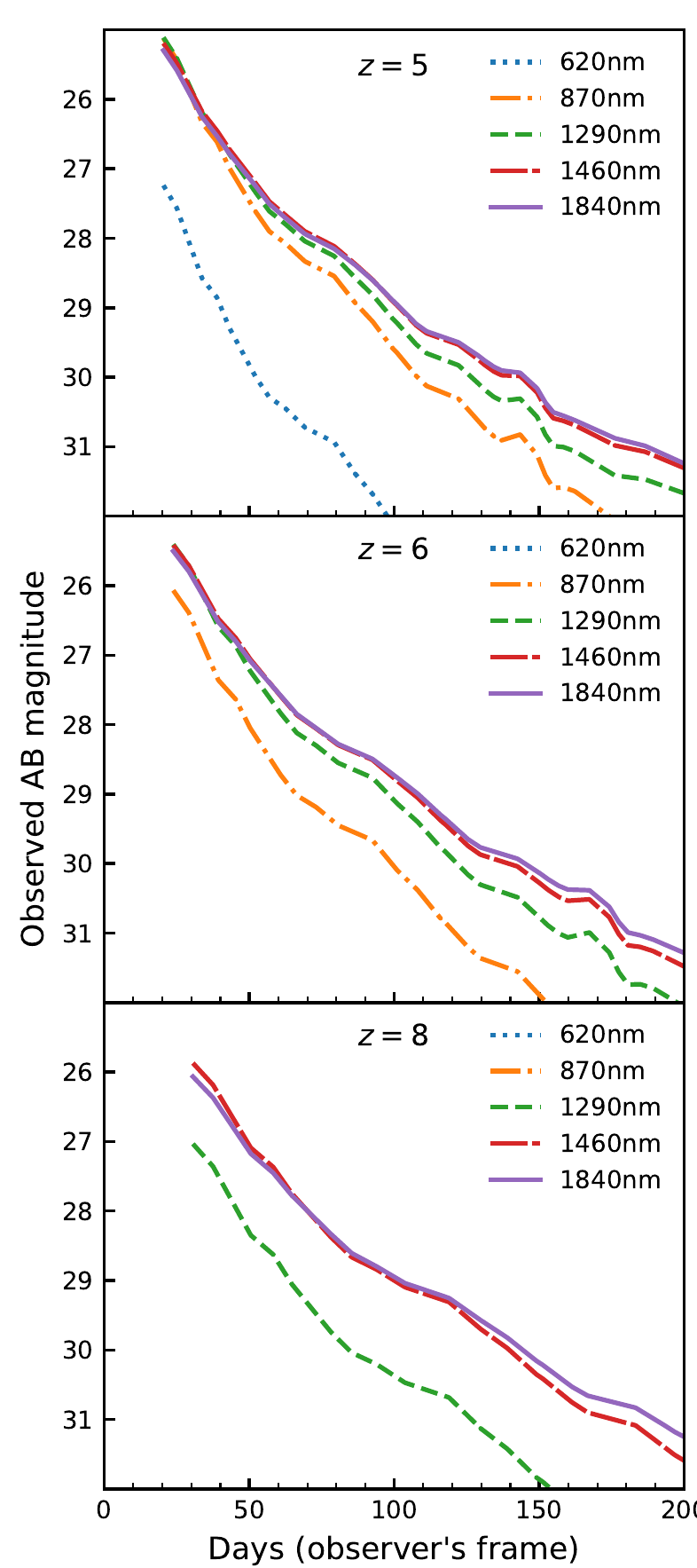}
    \caption{Observed-frame light curve of AT 2018cow model in each filters.}
    \label{fig:1}
\end{figure}

\begin{figure}
    \centering
    \includegraphics[width=\linewidth]{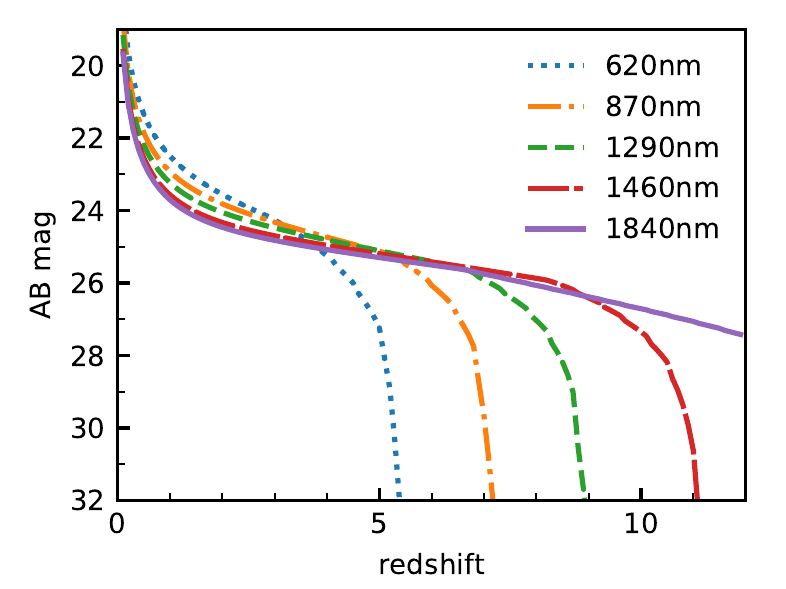}
    \caption{Peak magnitude of AT 2018cow model as a function of $z$.}
    \label{fig:2}
\end{figure}

We show light curves of the AT 2018cow model located at $z=5,6,8$ in Figure \ref{fig:1} and the peak magnitude of the AT 2018cow model as a function of $z$ in Figure \ref{fig:2}. From these figures we can observe the cosmological effects in the light curve, namely the time dilation and redshift (or K-correction).

\subsection{Rates of FLBTs}
The rate of FLBTs at high redshifts is dependent on their origin.
In the case of AT 2018cow, from the dense CSM indicated from the radio emission, \citet{Margutti2019} claims that AT 2018cow originates from an explosion of a massive star rather than tidal disruption of a white dwarf by an intermediate-mass black hole suggested by \cite{Perley2019a} and \cite{Kuin09}.
Thus, we assume that the frequency of FLBTs is proportional to the cosmic star formation rate (SFR) density $\rho_{*}$.
We adopt a fit of SFR density estimated from UV and IR observations of high-redshift galaxies (equation (1) of \citealt{Madau2017}):
\begin{equation}
    \rho_{*}(z) = 0.01 \frac{(1+z)^{2.6}}{1+[(1+z)/3.2]^{6.2}}\, M_{\odot} \, \rm{yr^{-1}\, Mpc^{-3}}.
\end{equation}

Similarly to \citet{Tanaka2013} who calculated the rate of superluminous supernovae, the event rate is written as
\begin{equation}
    R_{\rm{FLBT}}(z) = f_{\rm{FLBT}}\  \rho_{*}(z)\ 
    \frac{\int_{M_{\rm{min,CCSNe}}}^{M_{\rm{max,CCSNe}}} \psi (M) dM}{\int_{M_{\rm{min}}}^{M_{\rm{max}}} M \psi (M) dM},
\end{equation}
where $M_{\rm{min,CCSNe}}=8\, M_{\odot}$ and $M_{\rm{max,CCSNe}}=40\, M_{\odot}$ are the minimum and the maximum mass of stars which can produce core-collapse supernovae, and $\psi(M)$ is the initial mass function (IMF) of \citet{Kroupa2001}:

\begin{equation}
    \psi(M) \propto \left\{ \begin{array}{ll}
    M^{-0.3} & (0.01M_{\odot}(=M_{\rm{min}})<M<0.08M_{\odot}), \\
    M^{-1.3} & (0.08M_{\odot}<M<0.5M_{\odot}),\\
    M^{-2.3} & (0.5M_{\odot}<M<100M_{\odot}(=M_{\rm{max}})).
    \end{array} \right.
\end{equation}

We assume that a fraction $f_{\rm{FLBT}}$ of core-collapse supernovae explode as FLBTs like AT 2018cow.
From PAN-STARRS1 observations, \citet{Drout2014} estimated the rate of 
rapidly-evolving and luminous transients to be $4$ -- $7\,\%$ of the core-collapse supernova rate. 
Some of the PAN-STARRS1 examples have similar features as AT 2018cow, but not all samples are as bright as AT 2018cow, or are expected to be of stellar origin. 
%On the other hand the IMF is also uncertain. For example theoretical studies (e.g. \citealt{Susa14, Hirano15}) and observations of Galactic metal-poor stars \citep{Komiya07,Komiya09} imply that there are many more high-mass stars in the early universe than calculated from the Kroupa IMF, which may lead to a higher event rate.
In this work we simply set $f_{\rm{FLBT}}=0.01$. The actual number of detections will of course scale with $f_{\rm{FLBT}}$.

\subsection{Observation parameters and criteria}
To estimate the detection counts of FLBTs, we set the following observational parameters: the survey area, the cadence, and the limiting magnitude.
The survey area is supposed to be observed periodically with cadence of $\Delta t$ and the observed limiting magnitude is denoted by $m_{\rm{lim}}$.

We judge whether FLBTs can be detected or not according to the following two criteria.
\begin{enumerate}
    \item Magnitudes are less than the limiting magnitude in more than two filters.
    \item At least in one filter, the magnitude is less than the limiting magnitude at two consecutive epochs.
\end{enumerate}

\section{Result} \label{sec:result}
\begin{figure}
    \centering
    \includegraphics[width=\linewidth]{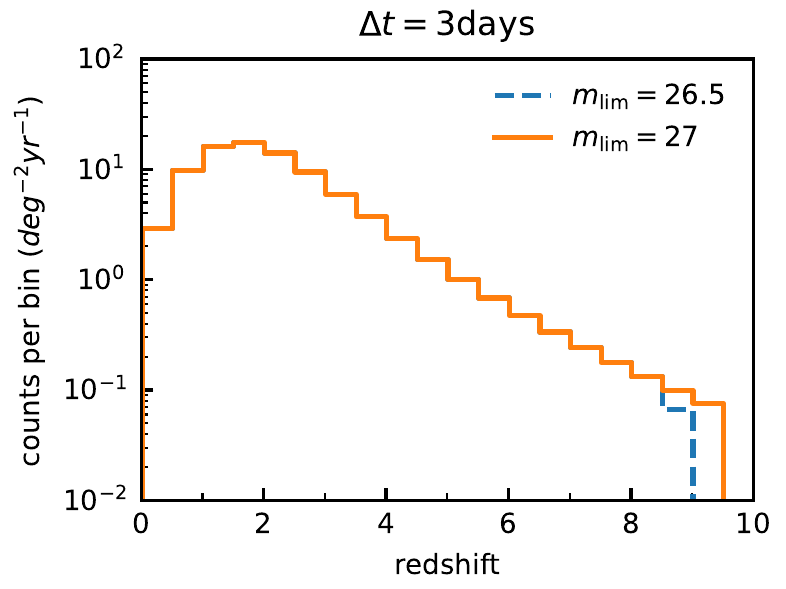}
    \caption{The detection counts per year per deg$^2$ per $dz = 0.5$ for a cadence of $\Delta t =3$ days. The two plots overlap except for $z>8$.}
    \label{fig:3}
\end{figure}

In Figure \ref{fig:3} we show the expected detection counts of FLBTs per year per deg$^2$ for $\Delta t = 3\,\rm{days}$, $m_{\rm{lim}}=26.5$ and $27$ mag.
We find that 2.1 (0.37) and 2.2 (0.48) FLBTs per year per deg$^2$ at $z>6$ ($z>8$) can be detected by each survey of $m_{\rm{lim}}=26.5$ and $27$.
It indicates that, to observe FLBTs in multiple bands using the present WFIRST filters, exposure time of $\sim10^3\,\rm{s}$ is sufficient \citep{Spergel2015}, and deeper observation can slightly extend the observable distance.

In contrast, the main reason why there are no detection beyond $z\sim9$ is that the magnitude at the F146 band suddenly decays at these wavelength regions, owing to the Lyman break (Figure \ref{fig:2}).
Therefore, observation for wavelengths longer than those considered in this work will be necessary to go beyond this redshift. Future telescopes that can do observations in e.g. K band may allow detection of more distant sources.

\begin{figure}
    \centering
    \includegraphics[width=\linewidth]{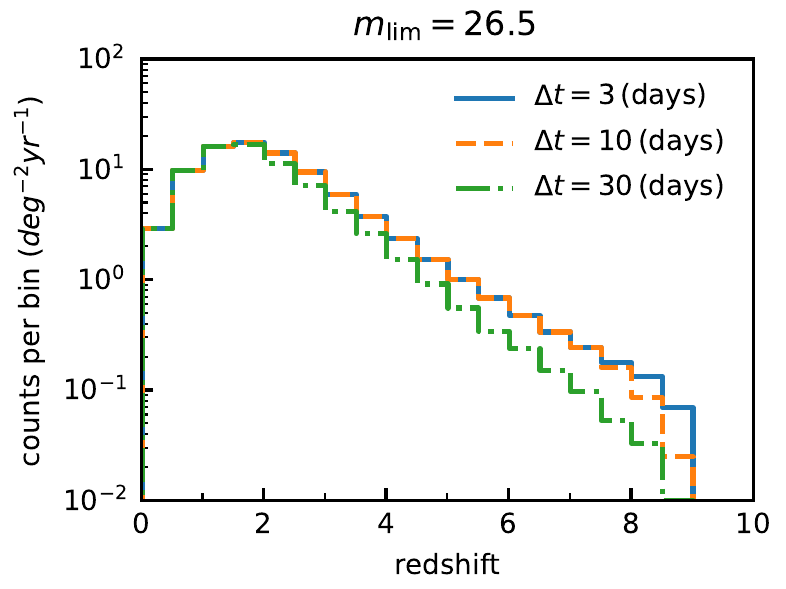}
    \caption{The detection counts per year per deg$^2$ per $dz = 0.5$ for $m_{\rm{lim}} = 26.5$ observation.}
    \label{fig:4}
\end{figure}

In Figure \ref{fig:4} we show the detection counts of FLBTs per year per deg$^2$ for $\Delta t = 3,\ 10,\ \rm{and}\ 30$ days.
We find that 2.1 (0.37), 2.0 (0.27), and 0.92 (0.096) FLBTs per year per deg$^2$ at $z>6$ ($z>8$) are detectable by each survey of $\Delta t = 3,\ 10,\ \rm{and}\ 30$ days.
Although the timescale of AT 2018cow is short, cosmic expansion dilates the duration of transients. Therefore, we do not require an extremely short cadence of observation.

From these estimates, we suggest that a survey with a limiting magnitude of $26.5$, a cadence of $\sim 10$ days, and with an area as wide as possible, will optimize the number of detectable FLBTs at high redshifts. For the upcoming WFIRST supernova deep survey, whose survey area is $\sim 5\ {\rm deg^2}$, $\Delta t = 5$ days, $m_{\rm lim}=26.5$, and survey period of $2$ years \citep{Hounsell18}, we predict that it is possible to detect $\sim 20$ $(\sim 4)$ AT 2018cow-like FLBTs at $z>6$ $(z>8)$.

\section{Discussion and Conclusion} \label{sec:discussion}
In this work we considered the detectability of AT 2018cow like transients at high redshifts, including the era of reionization.
We found that, if we assume the Kroupa IMF and SFR density inferred from galaxy observations, $\sim 20$ ($\sim 4$) FLBTs at $z>6$ ($z>8$) can be detected by WFIRST supernova (deep) survey with a limiting magnitude of 26.5 mag and a cadence of 5 days.

There will be some concerns when surveys are conducted in practice. First, to investigate the reionization, the FLBTs should be observed spectroscopically. To estimate the feasibility of spectroscopy, we refer to the observation of GRB 050904. \citet{Kawai06} obtained a clear spectrum of the afterglow with signal to noise ratio ($S/N$) $\sim5$ by a 4 hour spectroscopic observation. This spectrum covers the wavelength range of $7000$--$10000 \, \mathrm{\AA}$ with a resolution of $R \sim 1000$ at $9000\, \mathrm{\AA}$. \citet{Totani2006} analyzed this spectrum to obtain a stringent constraint on the degree of ionization in the IGM. 
We estimated the feasibility of similar spectroscopy for FLBTs at high redshifts with the Thirty Meter Telescope (TMT), by the InfraRed Imaging Spectrograph (IRIS) Exposure Time Calculator\footnote{\url{https://www.tmt.org/etc/iris}}. By a 5 hour observation ($R=4000$, wavelength range: $8400$--$10260\,\mathrm{\AA}$ (z band)) of a point source whose magnitude is 26.0 mag, the signal to noise ratio is estimated to be $S/N\sim5$. The required time can be reduced by a factor of 4 if the spectrum is binned to a coarser resolution of $R=1000$. Thus we conclude that spectroscopy is feasible for FLBTs using future thirty-meter class telescopes.

\begin{figure}[h]
    \centering
    \includegraphics[width=\linewidth]{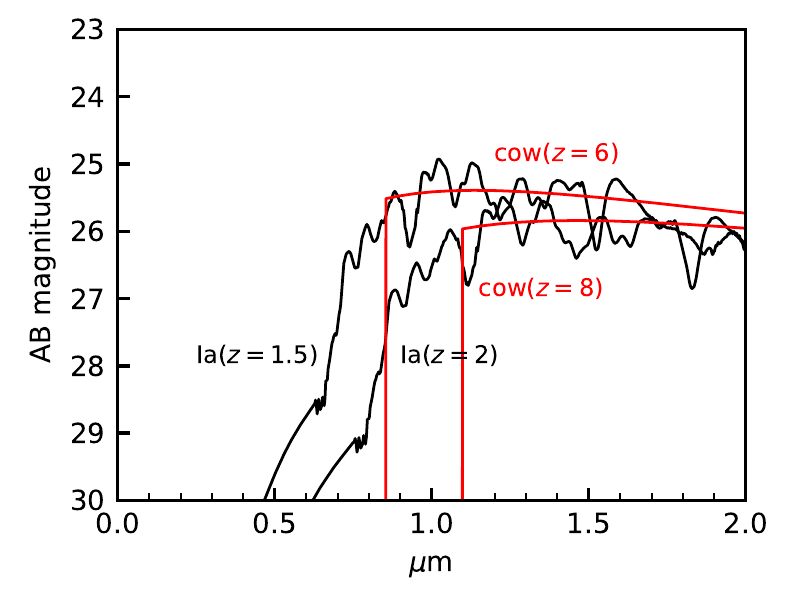}
    \caption{SEDs of AT 2018cow model at $z=6,\,8$ and  SED templates of Type Ia SN at $z=1.5,\,2$.}
    \label{fig:5}
\end{figure}

Second, FLBTs at high redshifts have to be distinguished from other transients at low redshifts, such as type Ia supernovae.
In this respect, the blueness of FLBTs makes the distinction between FLBTs at high redshift and other transients easy.
Figure \ref{fig:5} shows the SEDs of FLBTs at high redshifts and type Ia SNe at low redshifts whose magtitudes are comparable.
We use the type Ia SN template of \citet{Nugent2002} \footnote{\url{https://c3.lbl.gov/nugent/nugent_templates.html}}.
Because of the Ly$\alpha$ forest, the edges of the continuum of FLBTs become redder than those of type Ia SNe.
Thus, similarly to how high-redshift galaxies are selected according to Lyman breaks, high-redshift FLBTs can be distinguished from other low redshift transients with multi-wavelength data.
In practice, because both FLBTs and type Ia SNe have variation in luminosity and SED, the distinction according to photometric data may not be conclusive.
A follow-up observation to find the host galaxy after the transient has faded away can be another way of identifying its redshift.

Third, to constrain the neutral fraction of IGM, the degeneracy between extinction by neutral hydrogen in the host galaxy and that in the IGM has to be broken.
In the case of GRBs, \citet{Totani2006} combined the host galaxy absorption and the IGM absorption to fit the damping wing, and they set moderate constraints on the neutral fraction (however, see \citet{McQuinn2008}). In the FLBTs case, we expect that we can similarly conduct spectral fitting if the emission is a clean blackbody shape like AT 2018cow.
\cite{Totani2006} claimed that GRB 050904 was located in a host galaxy whose column density is high ($N_{\rm{HI}} \sim 4.0 \times 10^{21}\ {\rm{cm^{-2}}}$) and GRB with a low column density ($N_{\rm{HI}} \lesssim 10^{20}\ {\rm{cm^{-2}}}$) would constrain $x_{\rm{HI}}$ more accurately.
Similarly, if a FLBT occurs in a host galaxy whose column density is low ($N_{\rm{HI}} \lesssim 10^{20}\ {\rm{cm^{-2}}}$), it would be easier to resolve the degeneracy. 
We note that an upper limit comparable to this ($N_{\rm{HI}} < 3\times 10^{20}\ {\rm cm^{-2}}$) was reported by X-ray observations of AT 2018cow \citep{Margutti2019}. Thus follow-up deep observations to identify the host galaxy would be helpful if a viable candidate is found.

We conclude that FLBTs have sufficient potential as a probe for the epoch of reionization, and the upcoming WFIRST supernova survey can reveal more detailed information of reionization.

\acknowledgments
The authors thank the anonymous referee for comments that greatly improved this manuscript. The authors thank Yutaro Kofuji, Takehiro Yoshioka, and Kana Moriwaki for many helpful discussions. DT is supported by the Advanced Leading Graduate Course for Photon Science (ALPS) at the University of Tokyo. This work is also supported by JSPS KAKENHI Grant Numbers 19J21578, 16H06341, 16K05287, 15H02082, MEXT, Japan.

\bibliographystyle{aasjournal}

\end{document}